\documentclass[11pt,twoside]{article}
\usepackage{CAGN2016}
\usepackage{graphicx}

\usepackage[T1]{fontenc} 

\usepackage{latexsym}
\usepackage{verbatim}

\begin{document}

\vskip 1.0cm
\markboth{W. J.~Maciel et al.}{Chemical abundances in photoionized nebulae}
\pagestyle{myheadings}
%
%
\vspace*{0.5cm}
\parindent 0pt{Invited Review}


\vspace*{0.5cm}
\title{Chemical abundances of photoionized nebulae in the Local Group}

\author{W. J. Maciel$^1$, R. D. D. Costa$^1$ and O. Cavichia$^2$}
\affil{$^1$IAG, University of S\~ao Paulo, Rua do Mat\~ao 1226, 05508-090,
S\~ao Paulo SP, Brazil\\
$^2$IFQ, Universidade Federal de Itajub\'a, Av. BPS, 1303, 37500-903, Itajub\'a MG, 
Brazil\\ }

\begin{abstract}
Photoionized nebulae comprise basically HII regions and planetary nebulae, and their 
abundances give important clues on the nucleosynthesis and chemical evolution of their 
host galaxies. There is presently a large amount of data on these objects, especially 
for the elements He and N, which are strongly affected by the evolution of intermediate 
mass stars, as well as O, Ne, S, and Ar, which are essentially synthesized in stars with 
larger masses. The abundances of these elements in several systems in the Local Group are 
discussed on the basis of distance-independent correlations.
\end{abstract}

\section{Introduction}

Planetary nebulae (PN) provide accurate abundances of  elements that are not significantly produced by their 
progenitor stars such as O, Ne, S, and Ar, as well as some elements for which the abundances
have been changed, such as He, N, and C. The former can be used to study the chemical evolution
of the host galaxies, and the latter can place constraints on the nucleosynthesis of intermediate mass stars.
Distance-independent correlations involving O, Ne, S, and Ar can then be compared with the corresponding
abundances of young objects, such as HII regions, Blue Compact Galaxies (BCG) and Emission Line 
Galaxies (ELG). In this paper we intend (i) to compare the abundances of HII regions and PN in different
galaxies of the Local Group in order to investigate the differences derived from the age and origin of 
these objects, (ii) compare the chemical evolution in these systems, and (iii) investigate to what extent 
the nucleosynthesis contributions from the progenitor stars affect the observed abundances in planetary 
nebulae. Section~\ref{data} describes the data used in this investigation, and Section~\ref{results} 
presents our results and discussion. Further details can be found in Maciel et al. (2017).

\section{The Data}
\label{data}

We have considered abundance data for PN and HII regions in the following objects: The Milky Way (MW), 
the Large Magellanic Cloud (LMC), the Small Magellanic Cloud (SMC), M31, M32, M33, M51, M81, M101, NGC 185, 
NGC 205, NGC 300, NGC 628, NGC 3109, NGC 5194, and the Sextans galaxy. The uncertainties are typically
of 0.2 to 0.3 dex for PN and 0.1 to 0.2 dex for HII regions. The PN sample includes over 1300 objects, 
while the HII region sample has over 900 objects, as shown in Table~1. We have preferably used the PN 
data obtained by our own group, but have also considered some recent abundance determinations from the 
literature, particularly from sources using a similar procedure as our group. For HII regions we have
preferably adopted abundances obtained from detailed electron temperatures, instead of the strong line 
method. Blue Compact Galaxies and Emission Line Galaxies have also been included, as they are essentially
low metallicity HII regions. The samples are large enough to compensate for the inhomogeneity of data, 
as a large number of sources has to be considered, since there is no complete homogeneous sample available. 
The complete list of sources for each object is given by Maciel et al. (2017).

\begin{table*}
\small
\caption[]{Total samples.}
\label{table1}
\begin{flushleft}
\begin{tabular}{lrlr}
\noalign{\smallskip}
\hline\noalign{\smallskip}
Planetary Nebulae       & Number & \ HII Regions             & Number \\
\hline\noalign{\smallskip}
Milky Way Disk          & 347    & \ Milky Way               & 216 \\
Milky Way Bulge         & 267    & \ Magellanic Clouds       & 35  \\
Milky Way               & 614    & \ Other Galaxies          & 325 \\
Magellanic Clouds       & 511    & \ BCG, ELG                & 360 \\
Total External Galaxies	& 704    & \ Total External Galaxies & 720 \\
TOTAL                   & 1318   & \ TOTAL                   & 936 \\
\noalign{\smallskip}
\hline
\end{tabular}
\end{flushleft}
\end{table*}

\section{Results and Discussion}
\label{results}

\subsection{O, Ne, S, and Ar}

\begin{figure}
   \begin{center}
   \includegraphics[angle=-90, width=5.5cm]{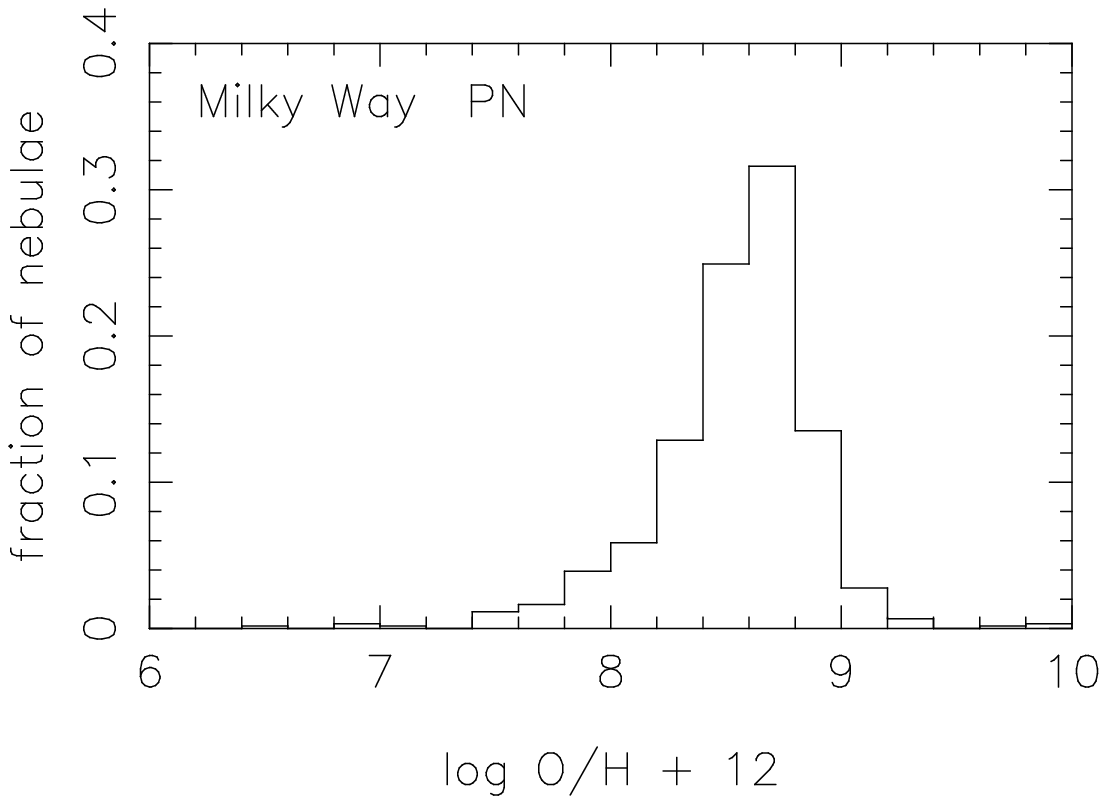}
   \includegraphics[angle=-90, width=5.5cm]{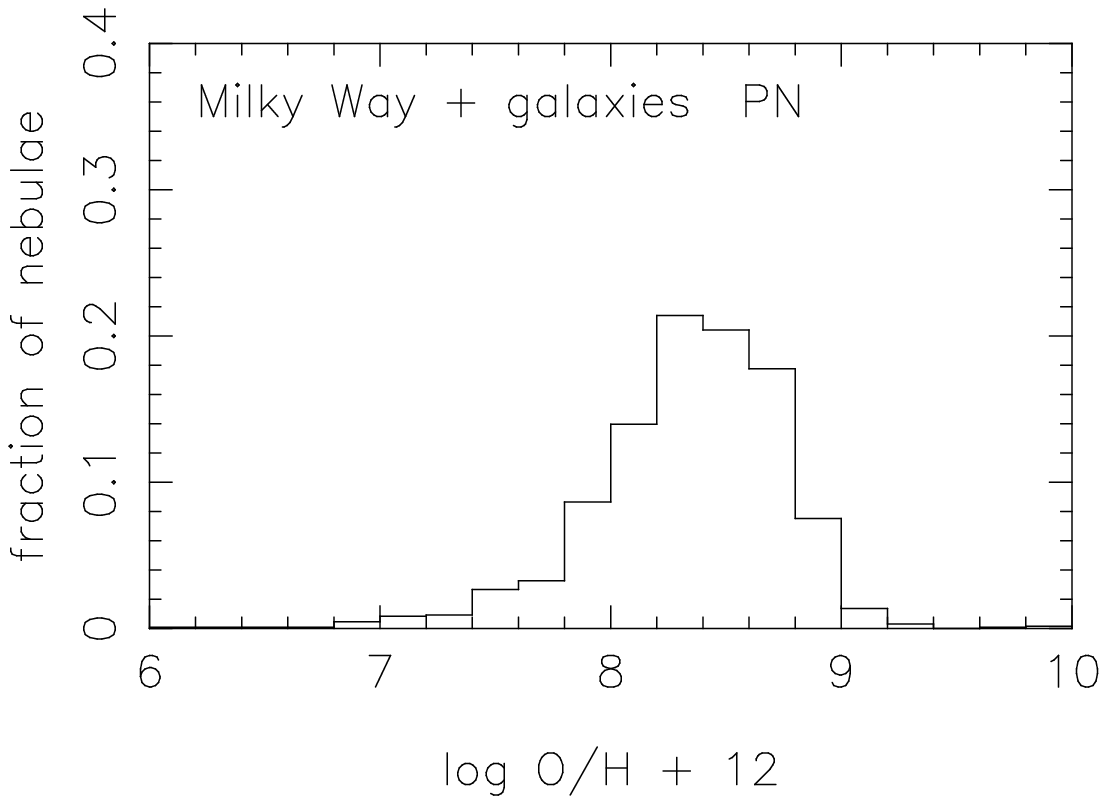}
   \includegraphics[angle=-90, width=5.5cm]{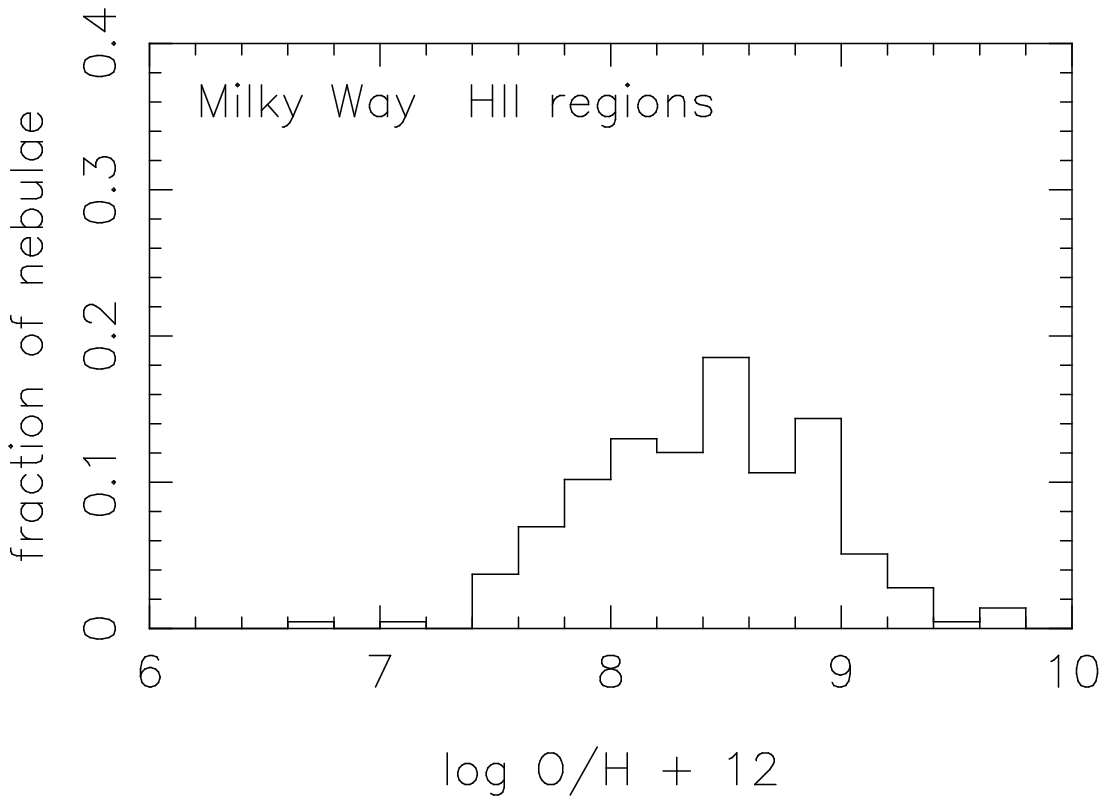}
   \includegraphics[angle=-90, width=5.5cm]{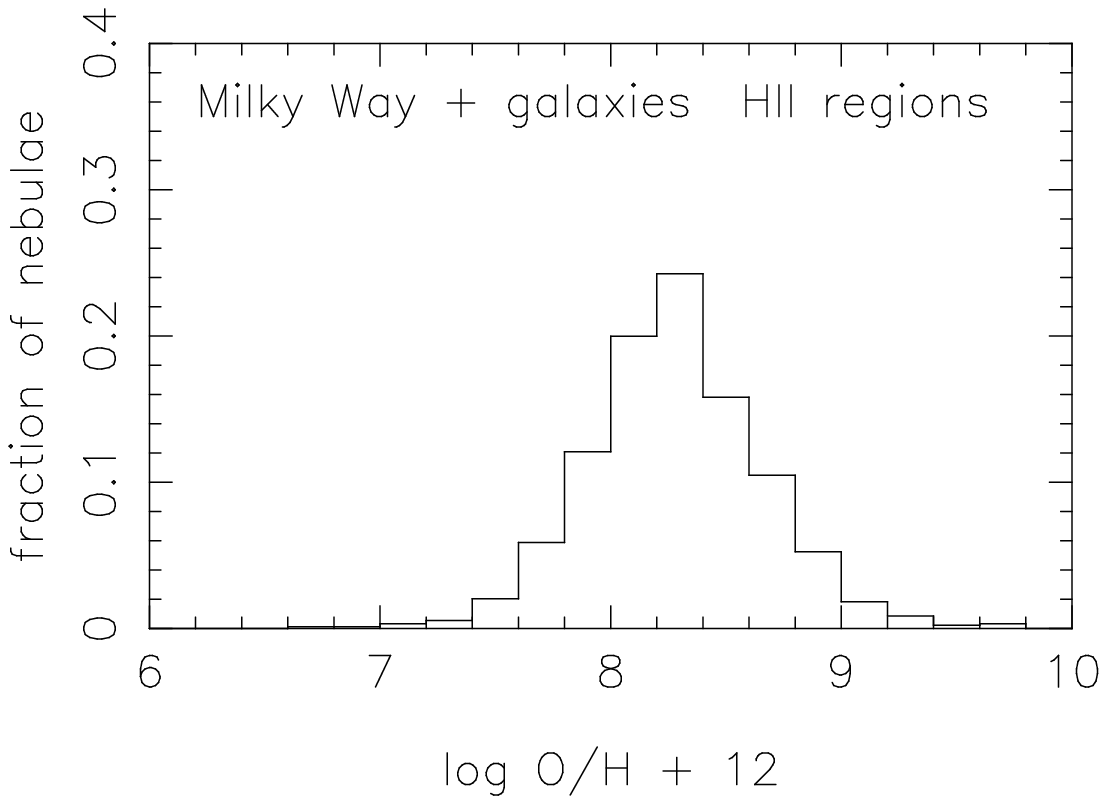}
   \caption{Histograms of the O/H abundances in PN and HII regions. Left: Milky Way;
   Right: Milky Way and external galaxies.}
   \label{fig1}
   \end{center}
   \end{figure}

Figure \ref{fig1} shows histograms of the oxygen abundance O/H for planetary nebulae and HII regions in
two representative cases: The Milky Way (left panels) and all objects considered here, namely the Milky 
Way, the Magellanic Clouds and the remaining external galaxies (right panels). Both PN and HII regions
have similar distributions, although the HII region distributions are generally broader than in the case 
of planetary nebulae. Also, it can be noticed that a larger fraction of HII regions have 
$\log {\rm (O/H)} + 12 \geq 9$, which reflects the fact that these younger objects are formed by more 
enriched material. Similar plots can be obtained for Ne, S, and Ar.

\bigskip

\begin{figure}
   \begin{center}
   \includegraphics[angle=0, width=5.5cm]{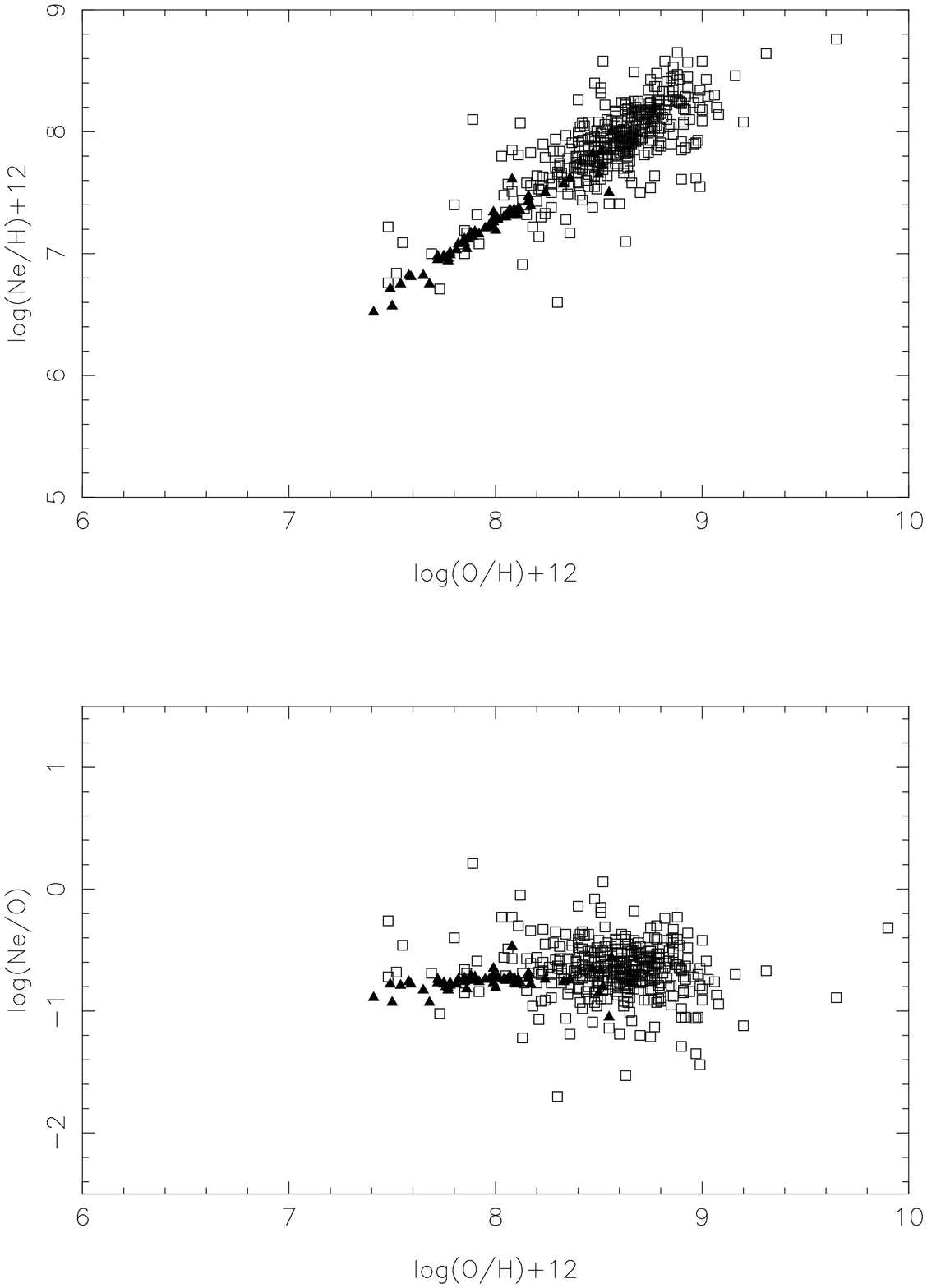}
   \includegraphics[angle=0, width=5.5cm]{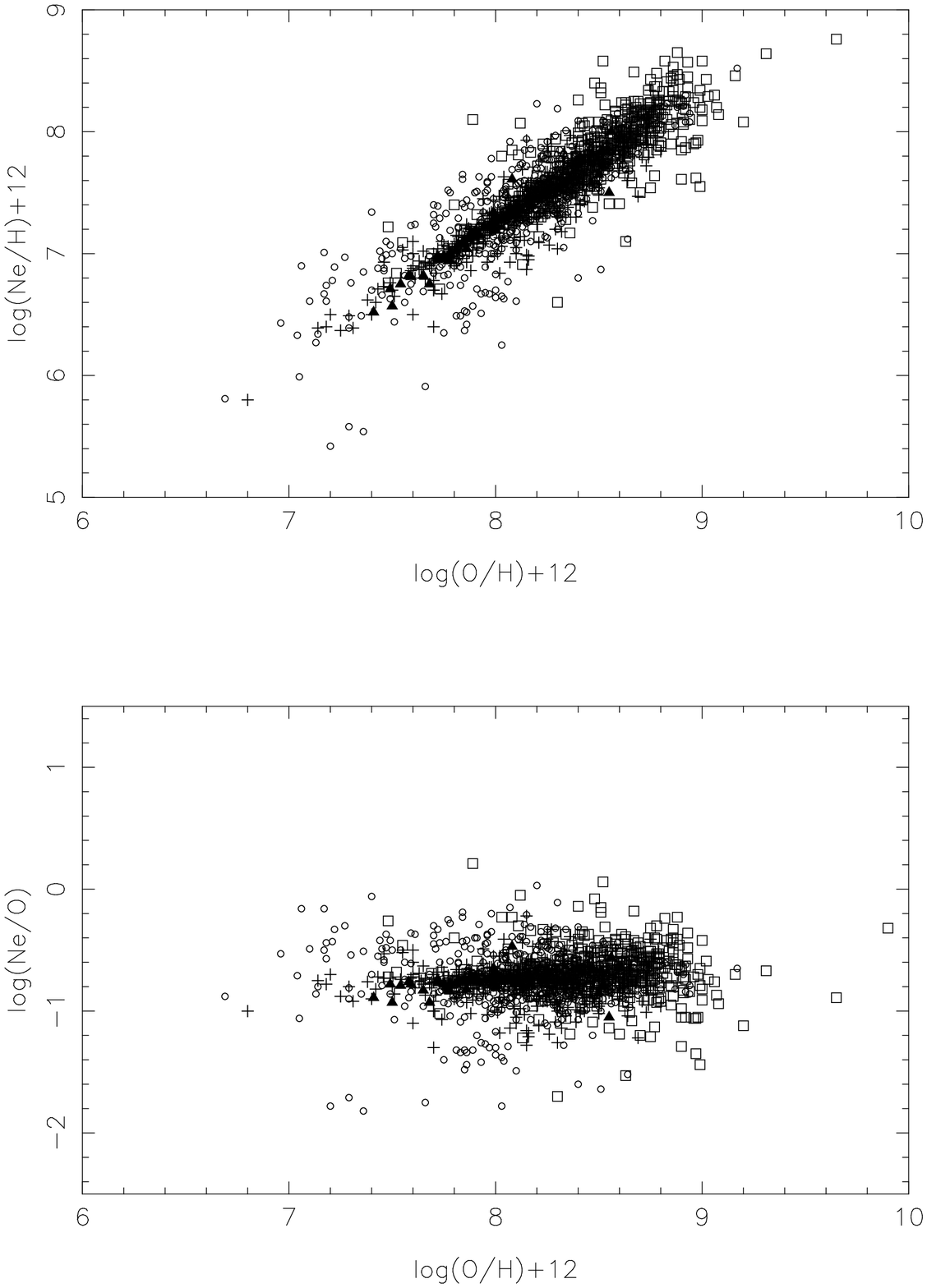}
   \caption{Ne abundances as functions of oxygen abundances. Left: Milky Way,
   right: Milky Way and external galaxies. Milky Way PN (squares), Milky Way  HII regions 
   (triangles); external PN: circles, external HII regions: crosses}
   \label{fig2}
   \end{center}
   \end{figure}

Distance-independent correlations for Ne are shown in Figure \ref{fig2}. The  left panels refer to the Milky Way, 
while the right panels are for all objects considered. The top figures show the abundances relative to 
hydrogen (Ne/H), while the bottom figures show the abundances relative to oxygen (Ne/O). 
The squares represent Milky Way PN, the circles are external PN, the triangles represent HII regions
in the Milky Way, and the crosses are external HII regions. We see that for the Ne/H ratio both PN and 
HII regions present a lockstep variation with O/H, although for HII regions the dispersion is much 
smaller. This is also reflected in the bottom figure, indicating that the Ne/O ratio is essentially 
constant with a higher dispersion for PN. The estimated dispersions are about 0.2 dex 
for PN and 0.1 dex for HII regions.

The same behaviour observed in the Galaxy also holds in other Local Group objects. Despite their 
different metallicities and morphologies, their nucleosynthetic processes and chemical 
evolution are apparently very similar. The trend displayed in Figure \ref{fig2} (right) shows a very 
good agreement with the trend found by Izotov et al. 2006 on the basis of ELG only. Similar conclusions 
were obtained by Richer and McCall 2007, 2008. For the right panel, the fractions of objects within 
1$\sigma$ and 2$\sigma$ are $0.78$  and $0.93$, respectively.

It is interesting to notice that in this larger sample the observed ranges of oxygen and neon abundances
in PN and HII regions are similar. The similarity essentially reflects the fact that the interstellar 
metallicities did not change appreciably in the last 5 Gyr approximately, a result that is supported by 
determinations of the age-metallicity in the Milky Way (see for example Rocha-Pinto et al. 2000, 
Bensby et al. 2004. 

The results shown in Figure \ref{fig2} can be interpreted assuming that the dispersion in the PN data reflects 
the fact that the abundances are not as well determined as in the HII regions. However, a larger dispersion 
would be expected, since PN are older objects than the HII regions and any given sample probably includes 
objects of different ages, as we have shown elsewhere (Maciel et al. 2010, 2011). 

It can also be considered that there is some contribution to the Ne abundances from the PN progenitor stars, 
as suggested in some investigations (see for example Pe\~na et al. 2007). It has been argued that  
the third dredge-up process in AGB stars may affect the oxygen abundances observed in PN 
(see for example Karakas \& Lattanzio 2014).  ON cycling would also reduce the O/H ratio especially in lower 
metallicity PN with massive progenitor stars (see for example Karakas \& Lattanzio 2007). Our results show that, 
if present, such contribution should be small compared with the average uncertainties in the PN abundances, 
so that an average contribution of about 0.1 dex cannot be ruled out. On the other hand, if we compare the 
expected contributions both to oxygen and neon, it is unlikely that their are equal, which is needed 
in order to explain the similarity of the PN and HII region trends shown in Figures \ref{fig2} 
(cf. Karakas and Lattanzio 2003). 

\bigskip

\begin{figure}
   \begin{center}
   \includegraphics[angle=0, width=6.5cm]{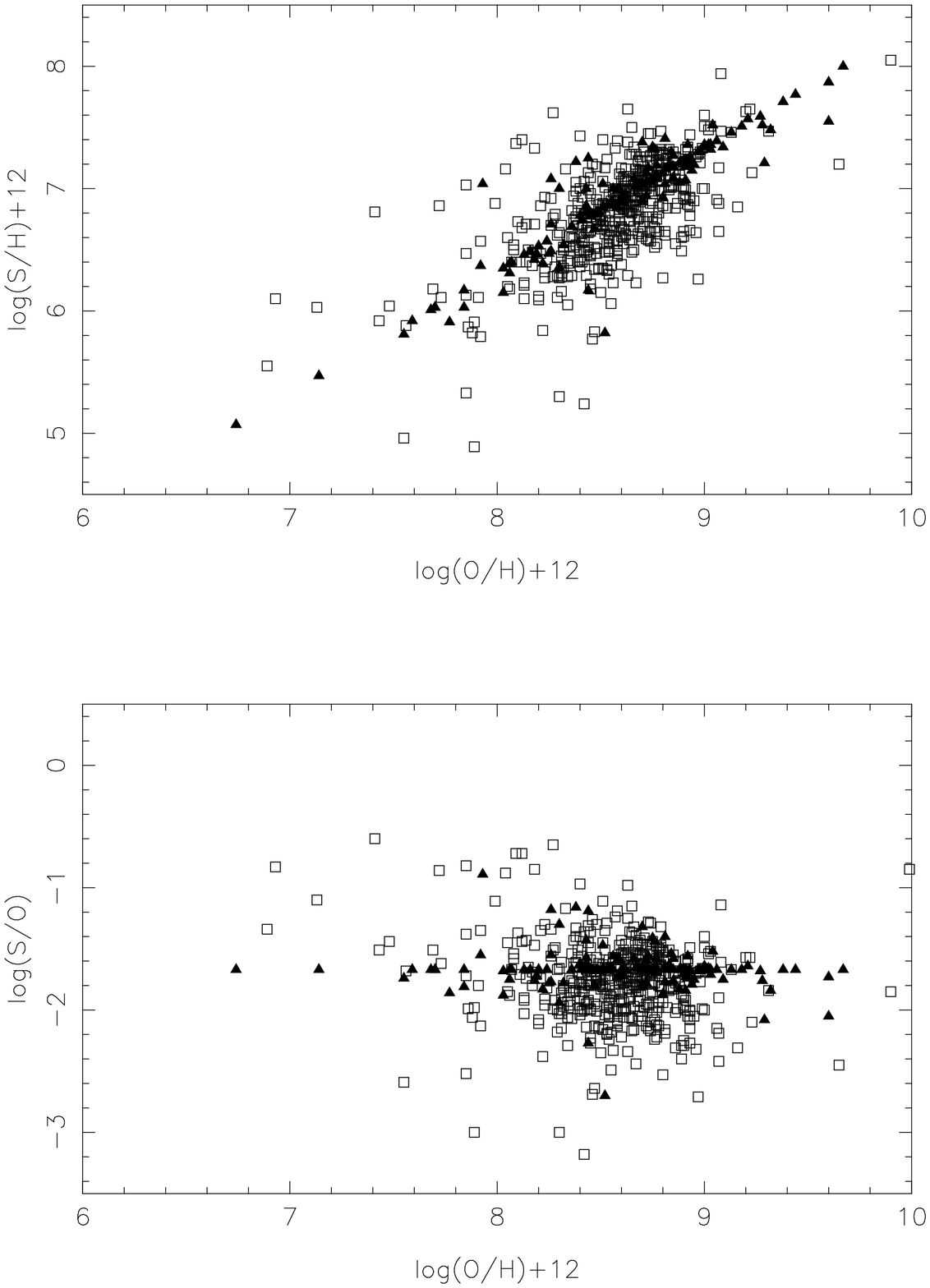}
   \includegraphics[angle=0, width=6.5cm]{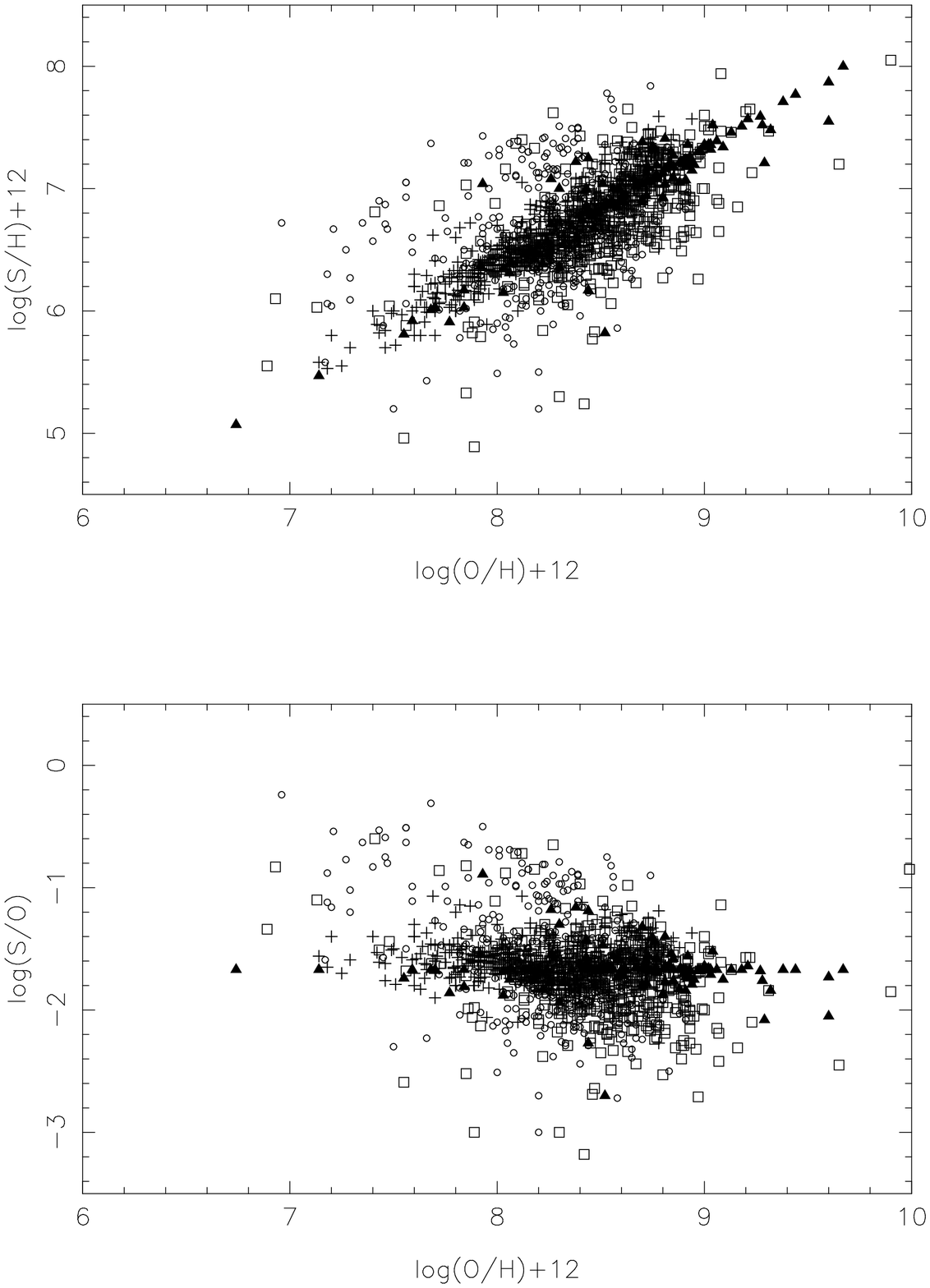}
   \caption{The same as Figure 2 for Sulphur}
   \label{fig3}
   \end{center}
   \end{figure}

In the case of sulphur, as shown in Figure \ref{fig3}, the general trends with oxygen are similar to neon, 
but some differences arise. The average dispersions are now 0.3 dex for PN and 0.2 dex for HII regions. 
For the Milky Way, the HII regions present a very good correlation, and the data extend to higher and lower 
metallicities compared with neon.  The galactic PN already display what is usually called the \lq\lq sulphur 
anomaly\rq\rq, that is, many PN apparently have somewhat lower S/H abundances than expected for their metallicity 
(see the detailed discussions by Henry et al. 2004, 2012). The sulphur anomaly has been attributed to a deficiency 
in the sulphur ICFs, particularly due to the abundance of the S$^{+3}$ ion, lack of accurate atomic constants, 
effect of the nucleosynthesis in the progenitor stars,  and different chemical evolution of the systems considered.  

Considering the right panel of Figure \ref{fig3}, it can be seen that there is still a reasonable number of 
objects below the HII region curve, but there is a large number of PN in the opposite side, so that the sulphur 
anomaly is not particularly noticeable.  The inclusion of BCG and ELG maintains these conclusions, that is, 
the sulphur abundances of HII regions apparently do not show the sulphur anomaly, which is then a characteristic of 
the empirical determination of sulphur abundances in PN. 

\bigskip

\begin{figure}
   \begin{center}
   \includegraphics[angle=0, width=6.5cm]{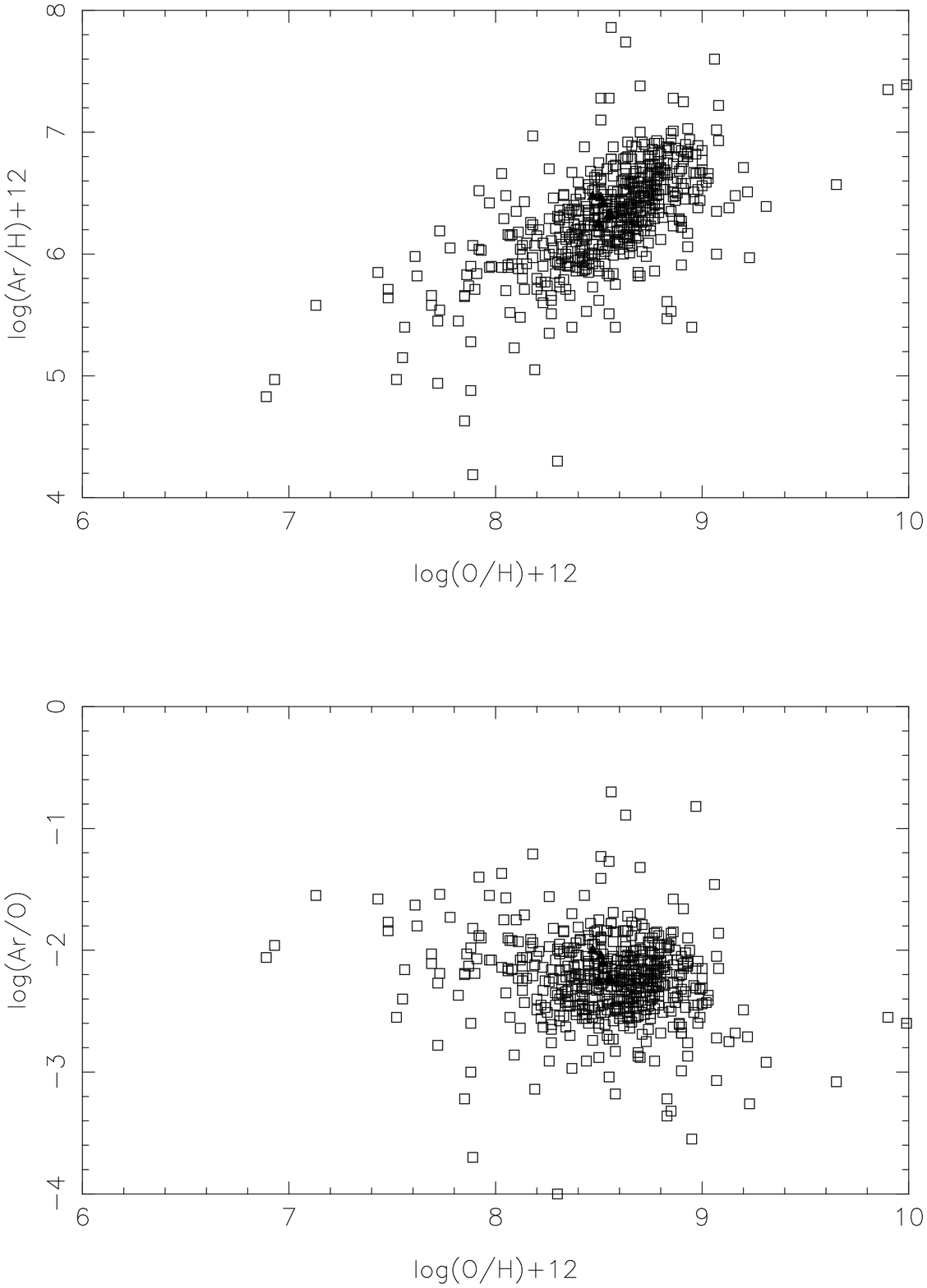}
   \includegraphics[angle=0, width=6.5cm]{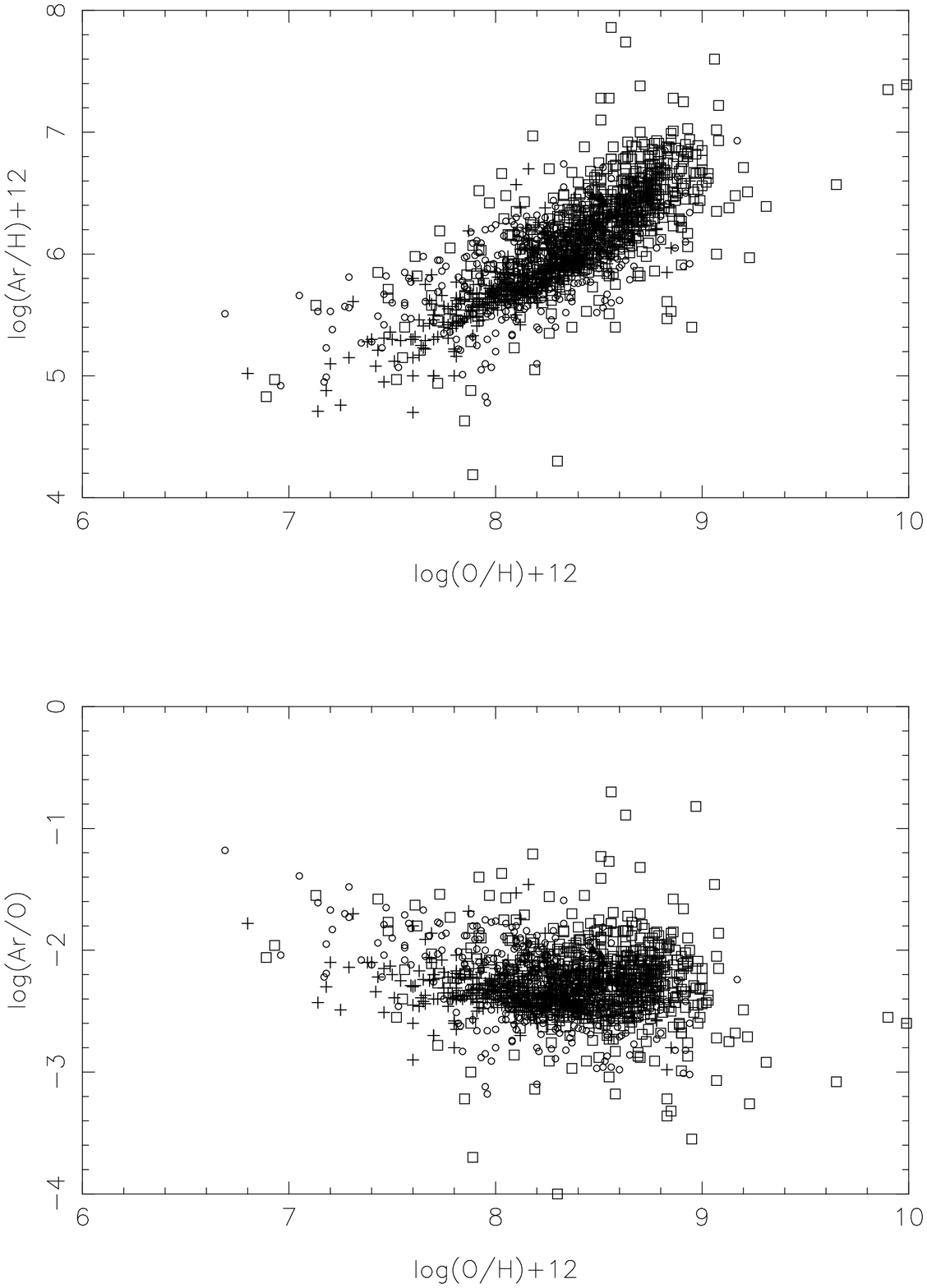}
   \caption{The same as Figure 2 for Argon}
   \label{fig4}
   \end{center}
   \end{figure}

For argon the observed correlations are similar compared to neon, but the dispersion is higher and is 
comparable to S/H, as can be seen from Figure \ref{fig4}.  The comparison with HII regions suffers from the 
lack of data for this element, especially for the Milky Way. The inclusion of BCG and ELG clearly improves 
the correlation, showing that the correlation defined at higher metallicities for the Milky Way still holds 
for lower and intermediate oxygen abundances. For PN, the Ar/H dispersion is higher compared 
with Ne/H, but for HII regions the dispersions in Ne/H, S/H, and Ar/H for the whole sample are similar to
each other, and always smaller than the PN data. Therefore, the HII regions are clearly more homogeneous than 
the PN, which reflects their very low ages, roughly a few million years. 

\subsection{Elements produced by the PN  progenitor stars}
\label{subsection32}

\begin{figure}
   \begin{center}
   \includegraphics[angle=-90, width=5.5cm]{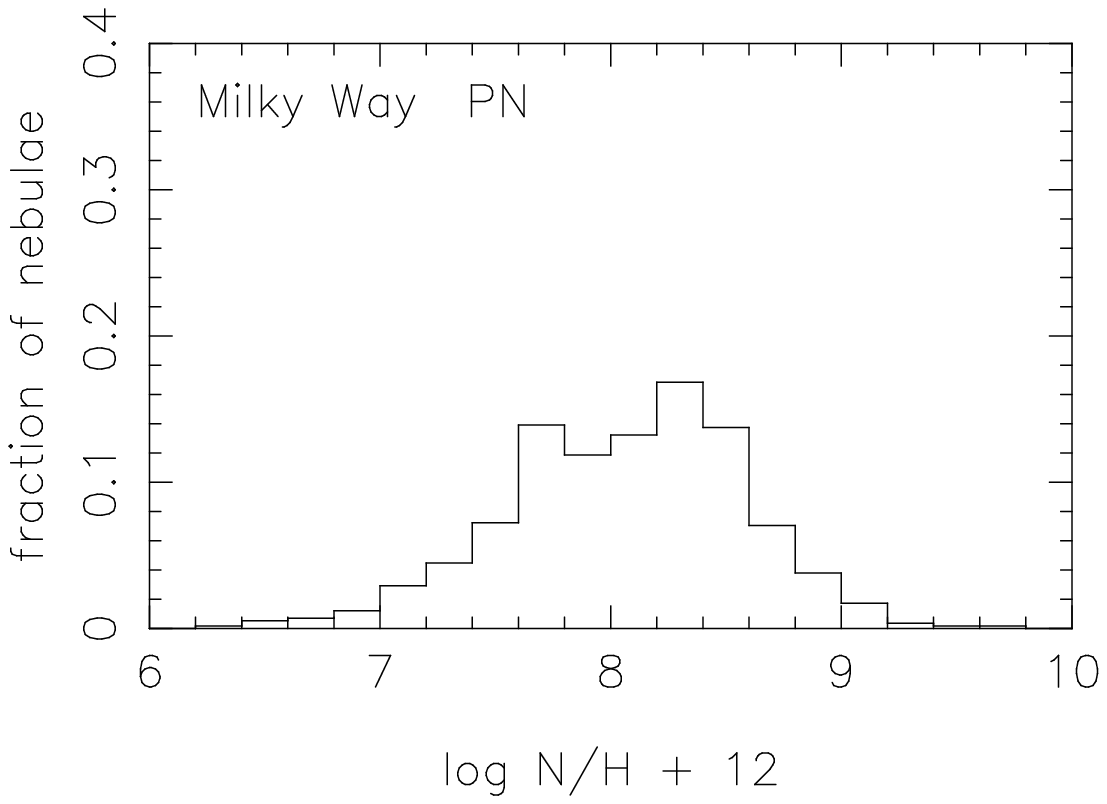}
   \includegraphics[angle=-90, width=5.5cm]{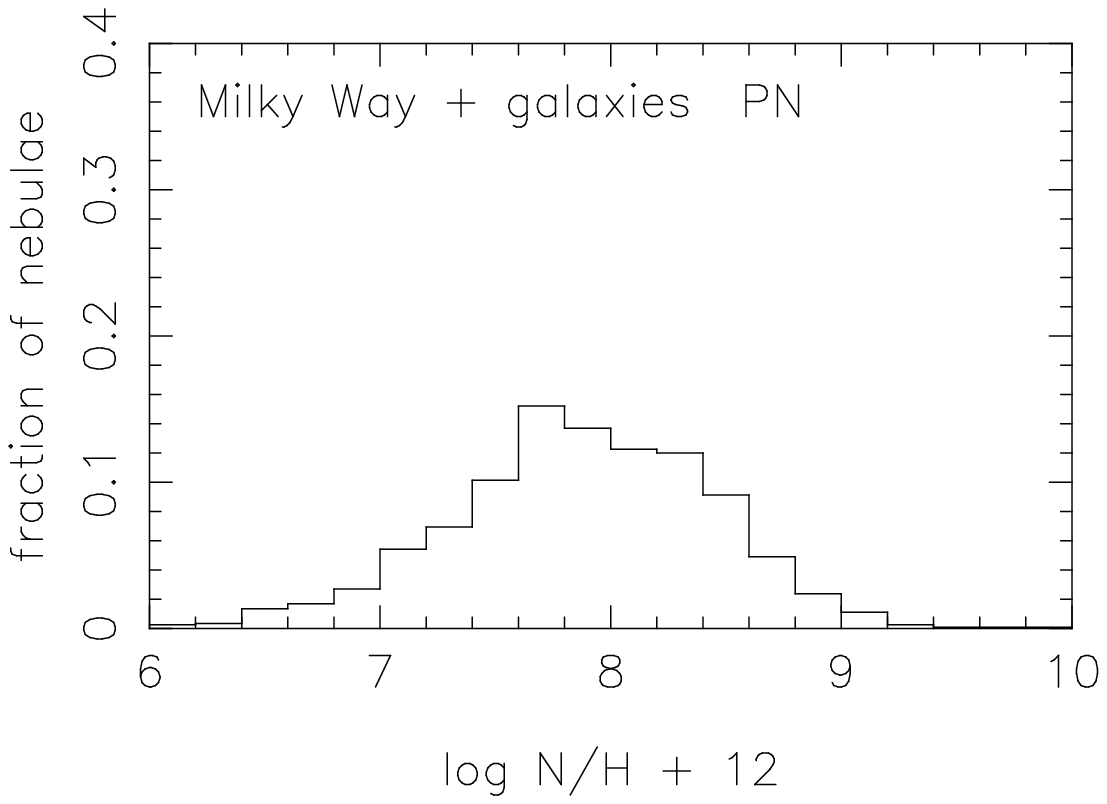}
   \includegraphics[angle=-90, width=5.5cm]{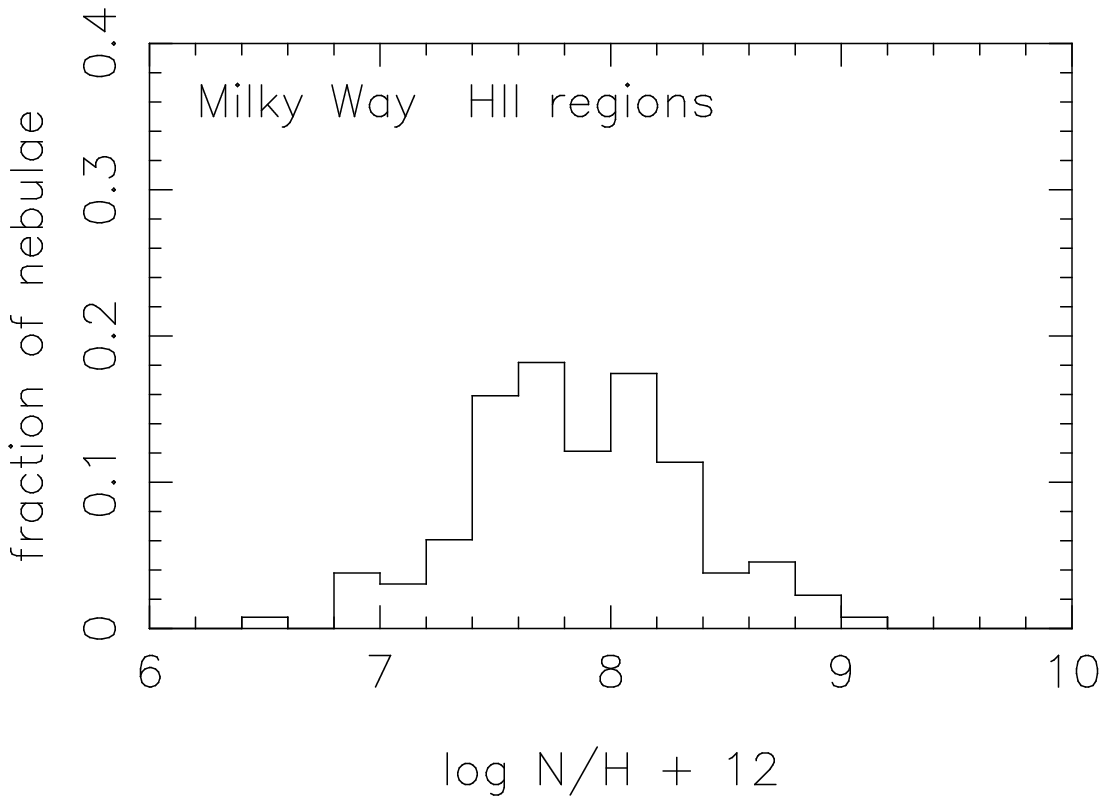}
   \includegraphics[angle=-90, width=5.5cm]{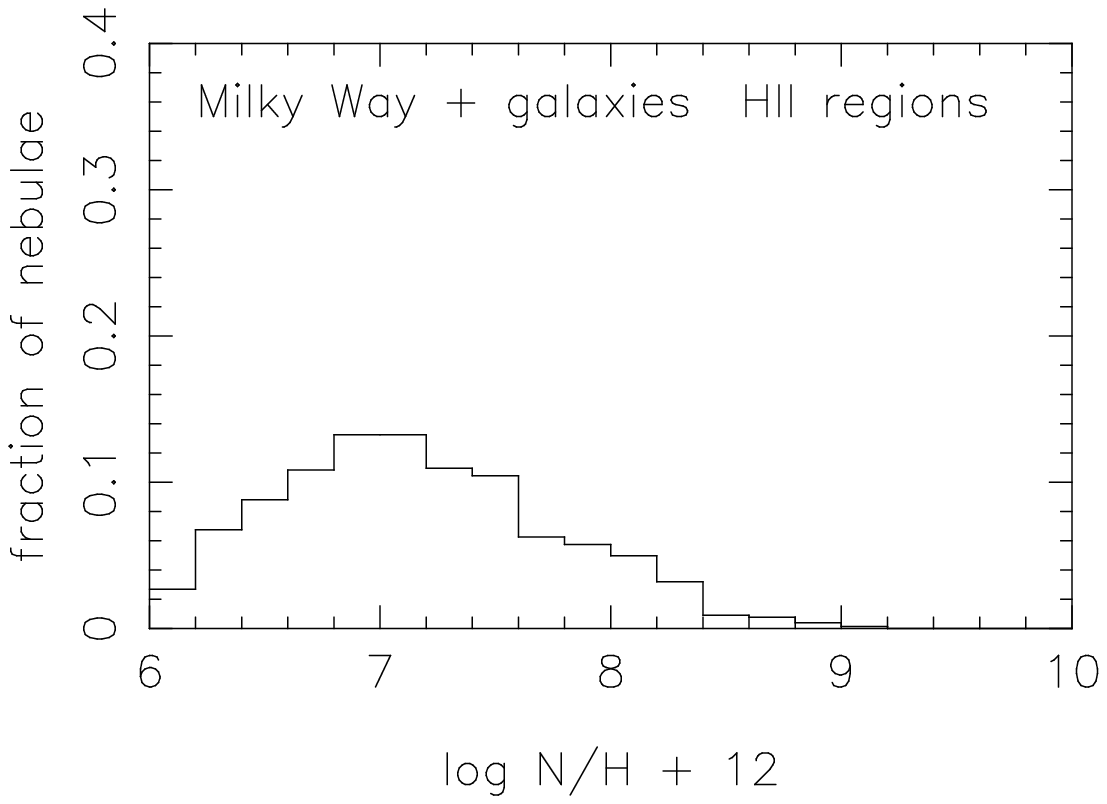}
   \caption{Histograms of the N/H abundances in PN and HII regions. Left: Milky Way,
    Right: Milky Way and all external galaxies.}
   \label{fig5}
   \end{center}
   \end{figure}

Histograms of the N/H abundances for PN and HII regions in our sample are shown in Figure \ref{fig5},
which can be directly compared with Figure \ref{fig1}. The PN distribution is similar in the two
cases shown, while for HII regions the inclusion of external galaxies (as well as BCG and ELG) shifts the 
maximum downwards by about 0.5 dex. The main difference between O/H and N/H is that the nitrogen abundances 
extends to lower metallicities for HII regions compared with PN, which reflects the N production 
during  the evolution of the PN progenitor stars.

\medskip

Helium abundances in HII regions  are frequently affected by the presence of neutral helium, so that in 
this work we have adopted a lower limit of He/H = 0.03 in order to avoid objects with an important 
fraction of neutral He.

   \begin{figure}
   \begin{center}
   \includegraphics[angle=0, height=6.0cm]{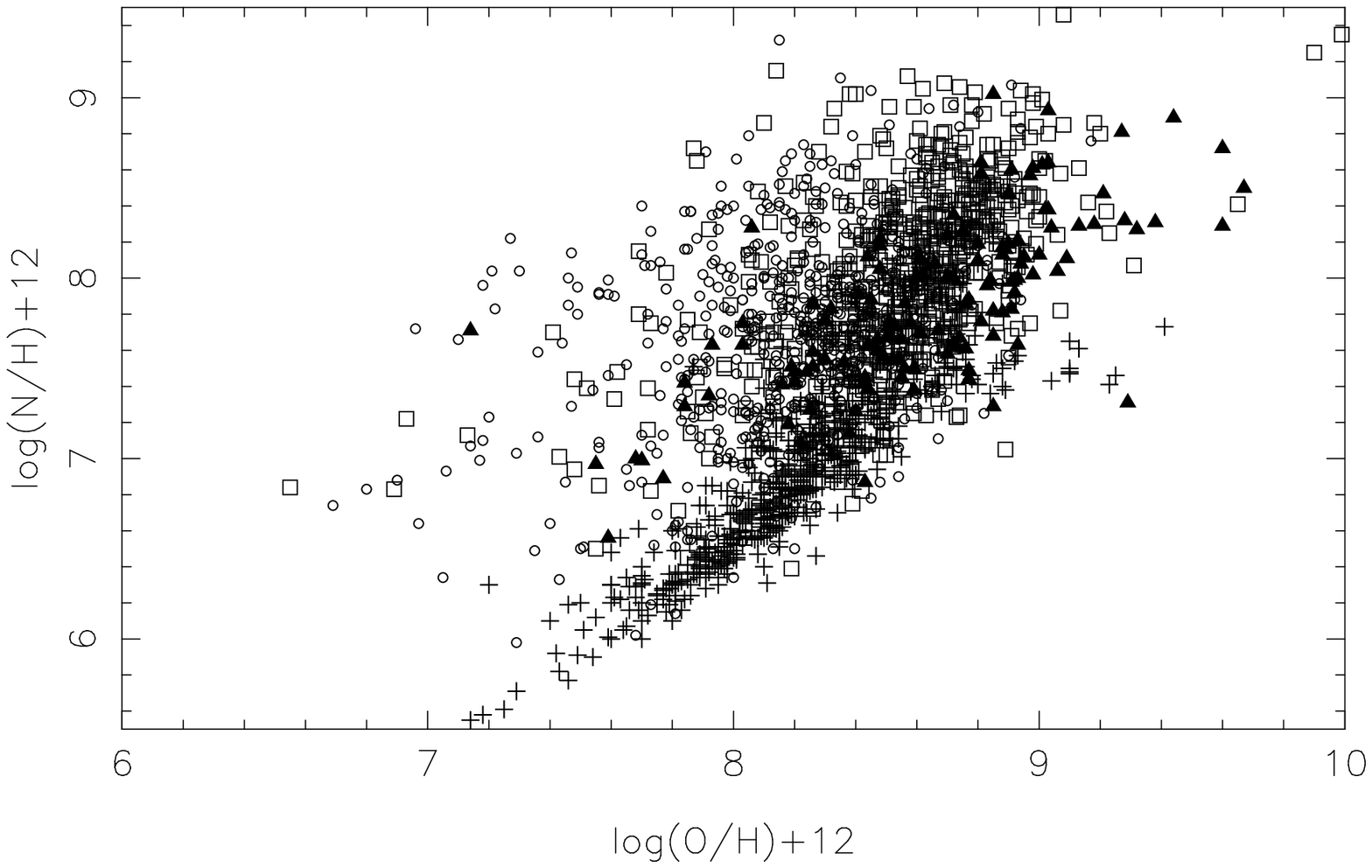}
   \includegraphics[angle=0, height=6.0cm]{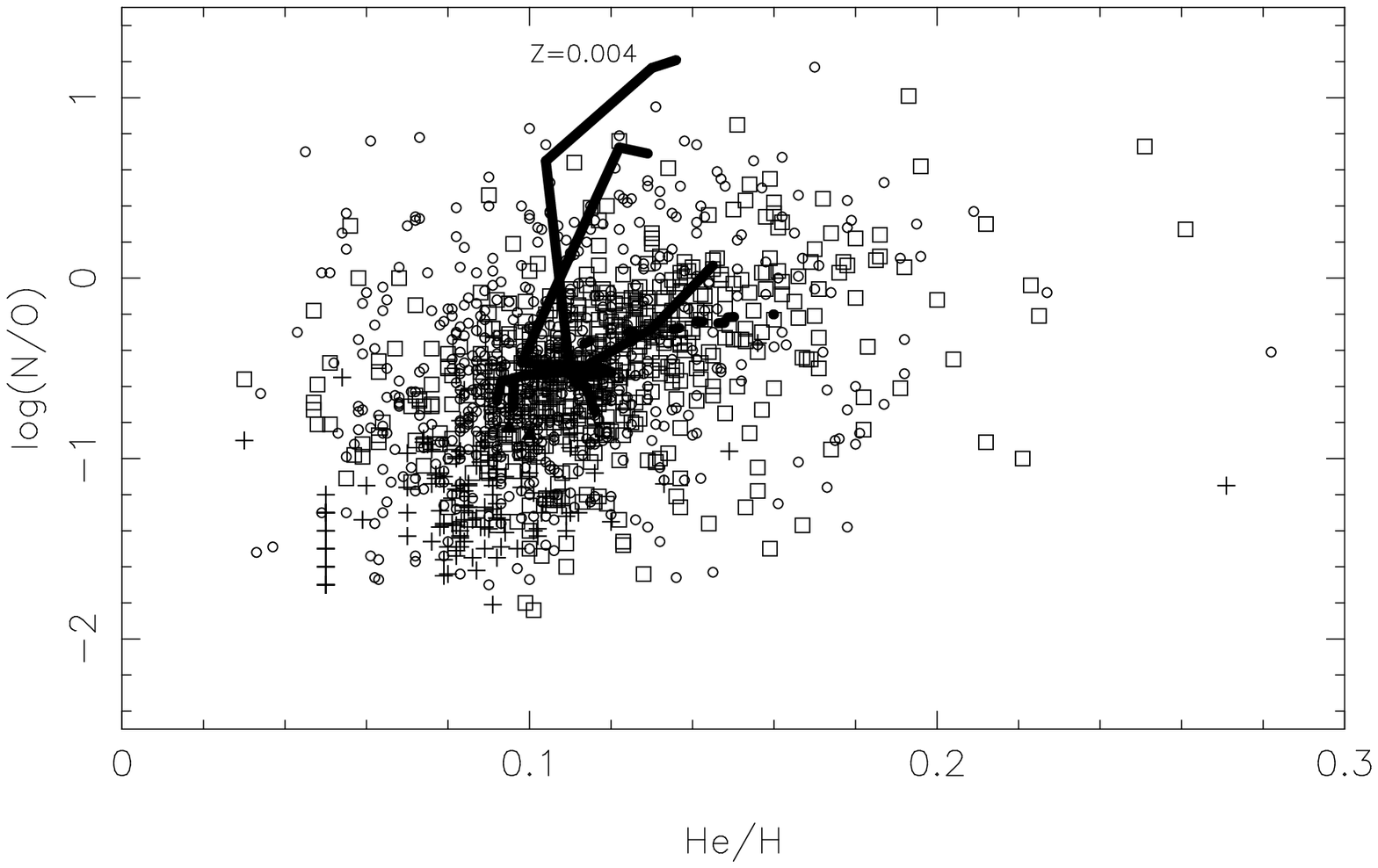}
   \caption{N abundances as functions of O/H (top) and  He/H (bottom) for the Milky Way and 
   external galaxies. MW PN (squares), MW HII regions (triangles), external PN (circles), 
   external HII regions (crosses). The lines are models by Karakas (solid lines) and Marigo 
   (dashed line).}
   \label{fig6}
   \end{center}
   \end{figure}

Figure \ref{fig6} shows the N/H and N/O ratios as functions of O/H and He/H, respectively,  for the case 
where all objects are considered. The most striking result is that, as expected, PN show an increase in both 
N and He compared to most HII regions in the sample.  The average dispersions of the nitrogen data are higher, 
about 0.4 dex for PN and 0.3 dex for HII regions. Therefore, a larger dispersion is also observed for HII 
regions, so that part of their nitrogen is probably secondary. 

Our results clearly reflect the fact that the nitrogen abundances measured in PN include both 
the pristine nitrogen plus the contribution from the dredge up processes that affect the red giant progenitor 
stars. Similar trends have also been recently discussed by Garc\'\i a-Hern\'andez et al. 2016.  Adopting a 
pregalactic He abundance by mass of about $Y = 0.255$ (cf. Izotov et al. 2014), which corresponds to 
approximately He/H = 0.09, we conclude from Figure \ref{fig6} (bottom) that about 80\% of the PN with He 
excess have abundances up to He/H $\simeq 0.141$, which is about 57\% higher than the pregalactic value. 
This can be compared with an amount of 50\% derived by Richer \& McCall 2008 from a smaller sample. More recently, 
Lattanzio \& Karakas 2016 suggested an increase of about 38\% in the helium content by mass from the second 
dredge-up process in AGB stars, which would lead to an increase of about 60\% in the He abundance by number  
of atoms, in excellent agreement with the results shown in Figure \ref{fig6}.

Also from Figure \ref{fig6} we can have an idea of the amount of nitrogen produced by the progenitor stars. 
Adopting as limit for primary nitrogen the amount produced by type II supernovae (Izotov et al. 2006), 
corresponding to approximately $\log {\rm N/O} = -1.6$, and considering the expected secondary nitrogen 
enrichment, which corresponds to about $\log {\rm N/O} = -1.2$,  Figure \ref{fig6} (bottom) implies that about 
80\% of the PN present an enrichment ratio up to  factor of 13.3, comparable with the factor of 10 found by 
Richer \& McCall (2016).

In Figure \ref{fig6} (bottom) we include a comparison with some recent theoretical models by  Karakas 2010 
with  $Z = 0.02,\ 0.004,$ \ and $0.008$, while the dashed lines represent models by Marigo et al. 2003 
with $Z = 0.019$. According to these models, progenitors having 0.9 to 4 $M_\odot$ and solar composition can 
explain the \lq\lq normal\rq\rq\ abundances, He/H $< 0.15$, while for objects with higher enhancements 
(He/H $ > 0.15$), masses of 4 to 5 $M_\odot$ are needed, plus an efficient HBB. Recent models by Pignatari et 
al. 2016 with $Z = 0.01$ and 0.02 are also consistent with these results, as can be seen in the discussion 
by Delgado-Inglada et al. 2015. For intermediate mass stars, agreement with theoretical models is fair, but 
abundance determinations should be improved and expanded.

\acknowledgments This work was partially supported by FAPESP, CNPq, and CAPES.

\end{document}